\documentclass[pra,twocolumn,showpacs]{revtex4}

\usepackage{amsfonts}
\usepackage{amssymb}
\usepackage[dvips]{graphicx}
\usepackage{amsmath}

\newcommand{\Tr}{\textrm{Tr}}

\newcommand{\JitterVariance}{\langle(\Delta \tau)^2\rangle_{\text{jitter}}}

\begin{document}

\title{Entanglement-based signature of nonlocal dispersion cancellation}
\author{Tomasz Wasak}
\affiliation{Faculty of Physics, University of Warsaw, ul.\ Ho\.{z}a 69, PL--00--681 Warszawa, Poland}

\author{Piotr Sza\'{n}kowski}
\affiliation{Faculty of Physics, University of Warsaw, ul.\ Ho\.{z}a 69, PL--00--681 Warszawa, Poland}

\author{Wojciech Wasilewski}
\affiliation{Faculty of Physics, University of Warsaw, ul.\ Ho\.{z}a 69, PL--00--681 Warszawa, Poland}

\author{Konrad Banaszek}
\affiliation{Faculty of Physics, University of Warsaw, ul.\ Ho\.{z}a 69, PL--00--681 Warszawa, Poland}

\begin{abstract}
We derive an inequality bounding the strength of temporal correlations for a pair of light beams prepared in a separable state and propagating through dispersive media with opposite signs of group velocity dispersion. The presented inequality can be violated by entangled states of light, such as photon pairs produced in spontaneous parametric down-conversion. Because the class of separable states covers the entire category of classical fields as a particular case, this result provides an unambiguously  quantum feature of nonlocal dispersion cancellation that cannot be reproduced within the classical theory of electromagnetic radiation.
\end{abstract}

\pacs{42.50.Xa, 03.65.Ud, 42.50.Nn, 03.67.Mn}
\maketitle

\section{Introduction}

The quantum theory of electromagnetic radiation allows for correlations substantially stronger than those permitted in classical models. Such correlations, resulting from the uniquely quantum phenomenon of entanglement, lead to striking effects in different degrees of freedom of light including polarization
\cite{KwiaMattPRL95}, wave vector \cite{PittShihPRA95}, and frequency. In the last case, one intriguing effect is the preservation of strong temporal correlations for fields traveling through dispersive media with opposite signs of the group velocity dispersion. This effect, known as nonlocal dispersion cancellation, has been described theoretically by Franson \cite{FranPRA92} and demonstrated in a recent experiment by Baek and coworkers \cite{BaekChoOPEX09}.

The actual role of quantum correlations in nonlocal dispersion cancellation has been a subject of a vexatious debate.
Various classical models have been presented to illustrate the gap between predictions of the classical and the quantum theories \cite{FranPRA92,BaekChoOPEX09,FranPRA09}. On the other hand, analogies between these two theories allow one to reproduce the effect of nonlocal dispersion cancellation using classical chaotic light up to certain features, such as appearance of a constant background in the detected signals \cite{TorrLajuNJP09,ShapPRA10,FranPRA10}. In this paper we present an inequality which bounds the strength of temporal correlations attainable when the fields are prepared in a separable state, i.e.\ when no quantum entanglement between the two light beams is present. This inequality, based on a separability condition for continuous-variable systems \cite{TanPRA99,BrauLoocRMP05}, shows that quantum entanglement is necessary to preserve all the properties of temporal correlations in nonlocal dispersion cancellation. At the same time, the presented result defines limits on how well classical fields can reproduce the effect of nonlocal dispersion cancellation. This is because the separability condition is satisfied by the
entire category of classical fields, which correspond in the quantum theory to statistical mixtures of coherent states  \cite{Prepresentation}, i.e.\ are represented by non-negative $P$-representations permitting only correlations of a separable character. Furthermore, the photodetection theory gives exactly the same predictions of photocount probabilities for classical fields and their quantum mechanical counterparts \cite{PhotonCounting}. Consequently, the derived inequality identifies a quantum feature of nonlocal dispersion cancellation that cannot be mimicked by the classical theory of optical radiation.

This paper is organized as follows. In Sec.~\ref{Sec:Propagation} we derive the transformation of the propagating fields using the Wigner phase space representation. The results are used to obtain the inequality in Sec.~\ref{Sec:Inequality}. Its application to a class of time-stationary Gaussian states is analyzed in Sec.~\ref{Sec:Gaussian}. Sec.~\ref{Sec:Experimental} discusses experimental aspects of testing the proposed inequality. Finally, Sec.~\ref{Sec:Conclusions} concludes the paper.

\section{Propagation}
\label{Sec:Propagation}

\begin{figure}
\centering
\includegraphics[scale=0.35,angle=-90]{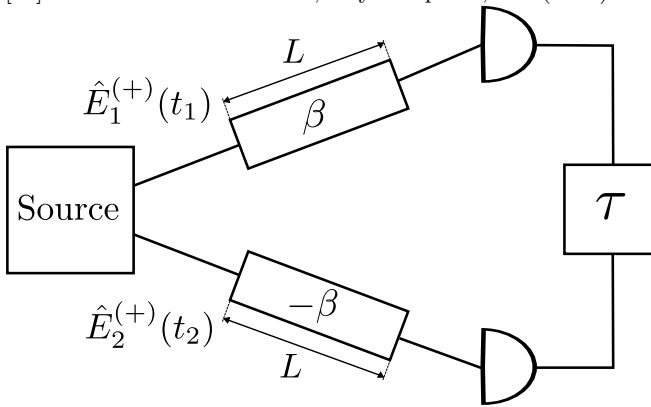}
\caption{\label{fig:system}
The measurement scheme. Two light beams characterized by positive-frequency field operators $\hat{E}_1^{(+)}(t_1)$ and $\hat{E}_2^{(+)}(t_2)$ propagate through dispersive media of equal lengths $L$ but opposite group velocity dispersions $\beta$ and $-\beta$. The quantity of interest is the time difference $\tau = t_1 - t_2$ between detection events.}
\end{figure}

The physical system under consideration shown in Fig.~\ref{fig:system} comprises two spatially separated light beams. We will describe the fields using annihilation parts of frequency-domain electric-field operators $\hat{\cal E}_j(\omega)$ parameterized with the detunings $\omega$ from the central frequency $\omega_0$ and satisfying commutation relations
\begin{equation}
\label{Eq:CommSpectr}
[\hat{\cal E}_i(\omega),\hat{\cal E}_j^\dagger(\omega')]=2\pi \delta_{ij} \delta(\omega-\omega'),
\end{equation}
where $i,j=1,2$.
We will assume that the fields under consideration have restricted bandwidth so that the slowly-varying envelope of the positive-frequency part of the electric field operator can be written in the temporal domain as:
\begin{equation}
\hat{E}^{(+)}_j (t) =
\frac{1}{2\pi} \int d\omega \, \hat{\cal E}_j(\omega)e^{-i \omega  t}.
\end{equation}

A transparent way to analyze the evolution of the system is to use the chronocyclic Wigner function
\cite{Chronocyclic} extended to a pair of optical beams:
\begin{widetext}
\begin{equation}
W(t_1, \omega_1; t_2, \omega_2) = \frac{1}{(2\pi)^2} \int d\tau_1
\int d\tau_2 \, e^{i\omega_1 \tau_1 + i \omega_2 \tau_2}
\Tr[\hat{\varrho}
\hat{E}_1^{(-)} (t_1 - {\textstyle\frac{\tau_1}{2}})
\hat{E}_2^{(-)} (t_2 -{\textstyle\frac{\tau_2}{2}})
\hat{E}_2^{(+)} (t_2 + {\textstyle\frac{\tau_2}{2}})
\hat{E}_1^{(+)} (t_1 + {\textstyle\frac{\tau_1}{2}})
],
\label{Eq:Wigner}
\end{equation}
\end{widetext}
where $\hat{\varrho}$ is the density matrix characterizing the fields and $\hat{E}_j^{(-)}(t)=[\hat{E}_j^{(+)}(t)]^\dagger$. In the regime of low light intensities, the integration of the Wigner function over frequencies $\omega_1$ and $\omega_2$ yields the joint probability of detecting photocounts at times $t_1$ and $t_2$, proportional to the expectation value
$\Tr[\hat{\varrho}\hat{E}_1^{(-)}(t_1)\hat{E}_2^{(-)}(t_2)\hat{E}_2^{(+)}(t_2)\hat{E}_1^{(+)}(t_1)]$.

Propagation through a dispersive medium of length $L$ results in multiplying the operators $\hat{\cal E}_j(\omega)$ by phase factors $e^{ik_j(\omega_0+\omega)L}$, where $k_j(\omega_0+\omega)$ is the wave vector at the frequency $\omega_0+\omega$ for the $j$th beam. We will apply the standard expansion up to the second order:
\begin{equation}
k_j(\omega_0+\omega) \approx k_j(\omega_0) + \frac{\omega}{v_j} + \beta_j \omega^2, \quad j=1,2
\end{equation}
where $v_j$ are group velocities and $\beta_j$ are parameters characterizing the group velocity dispersion. A straightforward calculation shows that in the Wigner representation the effect of dispersive propagation is given by the following transformation of the temporal variables:
\begin{equation}
\label{Eq:ttransf}
t_j' =  t_j + \frac{L}{v_j} + 2 \beta_j L\omega_j
\end{equation}
while the frequencies $\omega_j$ are unaffected. We use prime signs to denote the chronocyclic variables after propagation. It is worthwhile to note that the above result is analogous to the quantum mechanical evolution of a free particle \cite{IBBQM}.

\section{Inequality}
\label{Sec:Inequality}

\begin{figure}
\centering
\includegraphics[scale=0.25]{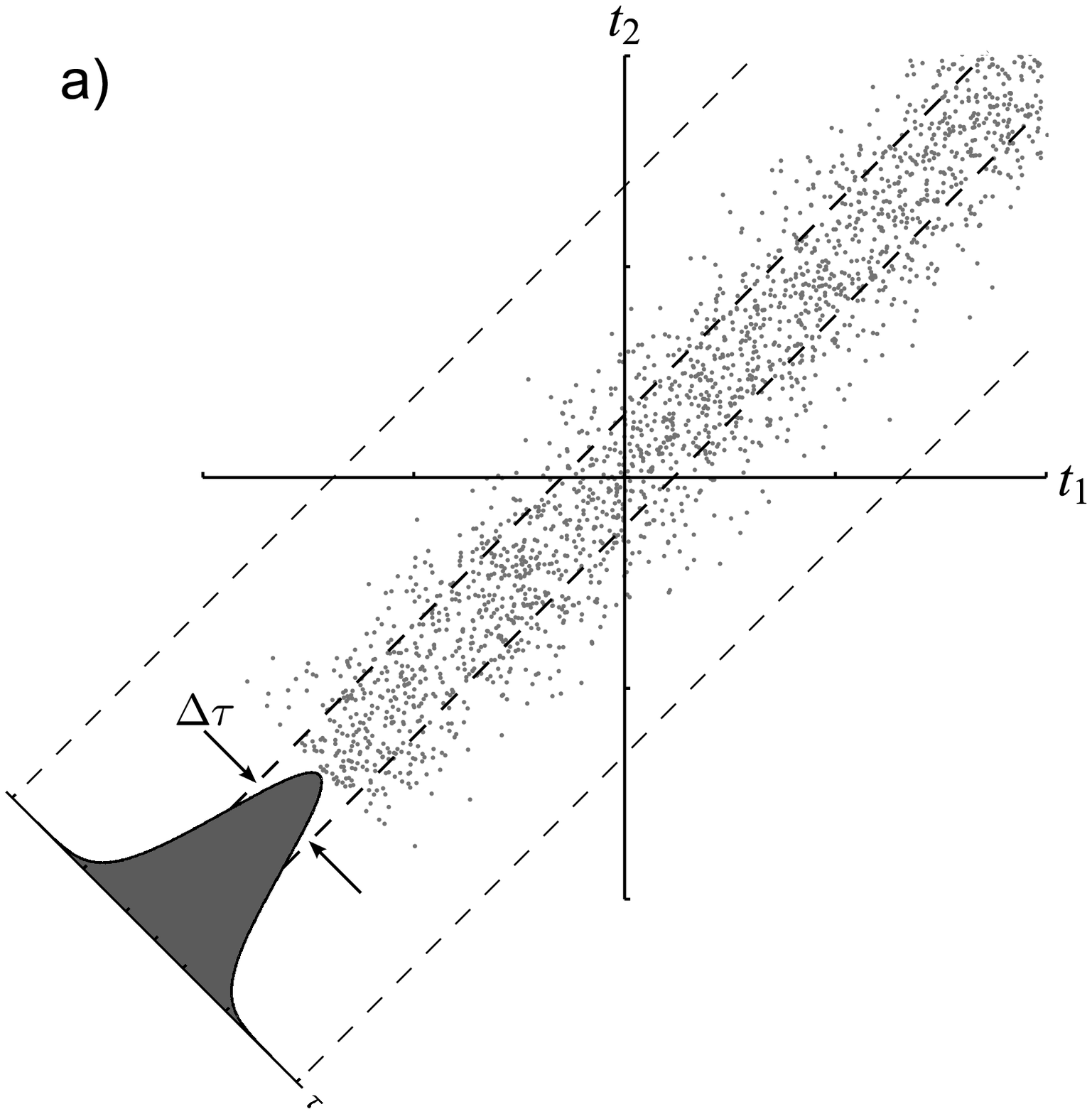}\includegraphics[scale=0.25]{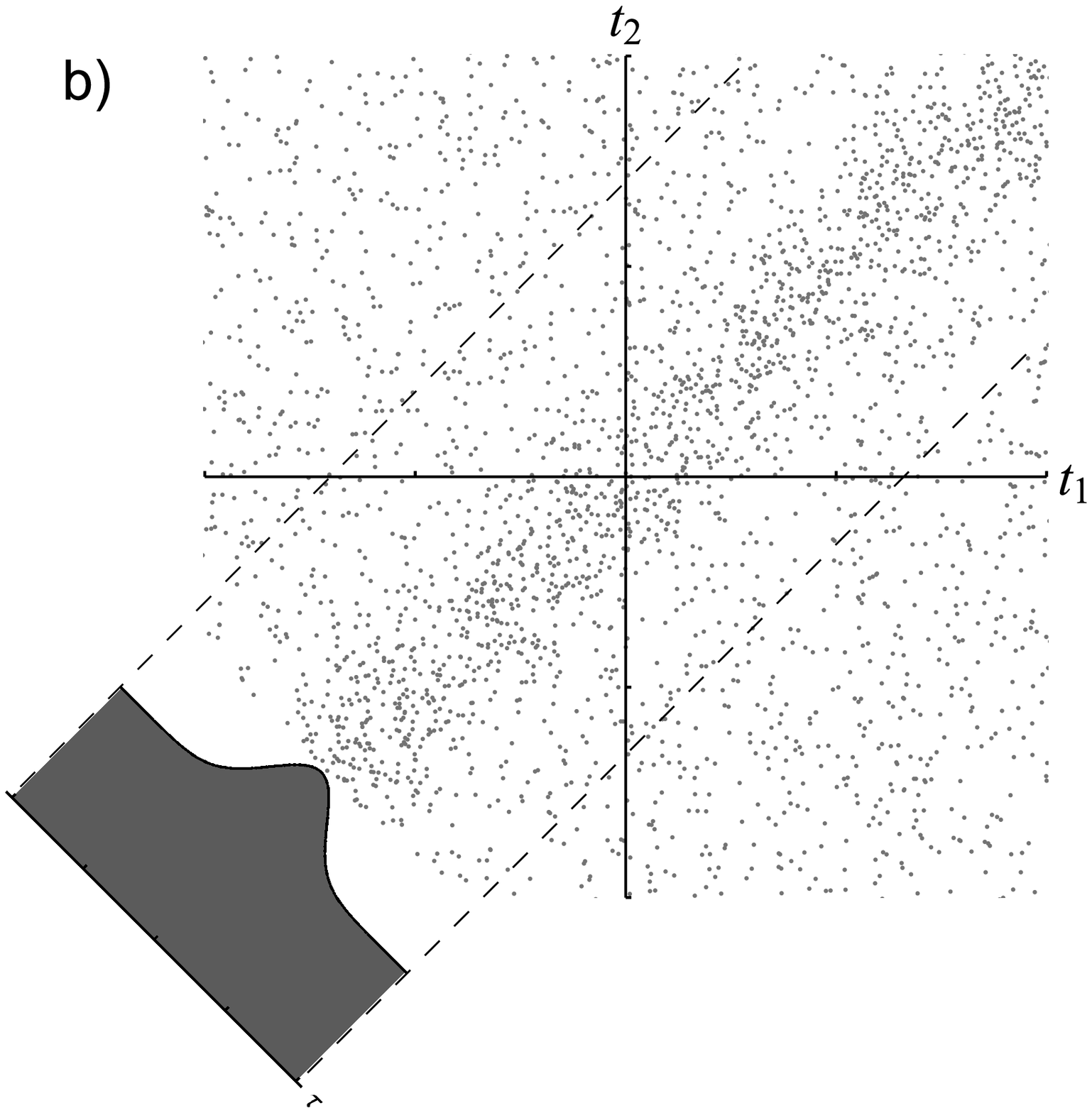}
\caption{Joint time statistics of detection events. Each dot $(t_1, t_2)$
represents registration of two photons at times $t_1$ and $t_2$. The object of interest is the marginal
distribution of the time difference $\tau=t_1-t_2$. In the quantum case (a) all the events
are temporally correlated, giving a finite value of the variance
$(\Delta\tau)^2$. Classically (b), the constant background makes the variance divergent.}
\label{Fig:whatistau}
\end{figure}

With Eq.~(\ref{Eq:ttransf}) at hand, the remaining analysis is elementary. Let us consider the time difference between the detection events $\tau = t_1 - t_2$, depicted schematically in Fig.~\ref{Fig:whatistau}(a). The statistical properties of this quantity are obtained by averaging with the normalized Wigner function, which we will denote by angular brackets $\langle \ldots \rangle$. For group velocity dispersion parameters of equal magnitude but opposite signs, $\beta_1 = -\beta_2 = \beta$, one immediately obtains that the time difference variance after propagation $\langle (\Delta \tau' )^2 \rangle$ is given by the following combination of the covariance matrix elements for the initial state of light:
\begin{equation}
\langle (\Delta \tau' )^2 \rangle = \langle (\Delta \tau)^2 \rangle +
 4 \beta L \langle \Delta\tau \Delta\Omega \rangle  +
 (2 \beta L)^2 \langle (\Delta \Omega)^2 \rangle
 \end{equation}
where $\Omega = \omega_1 + \omega_2$ is the sum frequency for both the photons. It is seen that the variance of the time difference is modified
by two contributions. The first one comes from the mixed time-frequency covariance $\langle \Delta\tau \Delta\Omega \rangle$. It can be eliminated by an assumption that the quantum statistical properties of the beams are invariant with respect to their physical interchange, which flips the sign of $\tau$. Alternatively, suppose that we perform two separate experiments with swapped positions of the dispersive media, and take an arithmetic average $\langle (\Delta \tau' )^2 \rangle_\text{sym}$ of the two variances measured after propagation. As the parameter $\beta$ has opposite signs in both the cases, this also removes the contribution from the mixed covariance term. Thus in the symmetric scenario, the time difference variance is simply enhanced by the variance of $\Omega$:
\begin{equation}
\label{Eq:Deltatausym}
\langle (\Delta \tau' )^2 \rangle_\text{sym} = \langle (\Delta \tau)^2 \rangle +
 (2 \beta L)^2 \langle (\Delta \Omega)^2 \rangle.
\end{equation}

The question about the role of entanglement in nonlocal dispersion cancellation now boils down to limitations that need to be satisfied by the variances $\langle (\Delta \tau)^2 \rangle$ and $\langle (\Delta \Omega)^2 \rangle$. From the formal point of view, our problem is analogous to that of a pair of continuous-variable systems, such as particles in one spatial dimension or single light modes \cite{BrauLoocRMP05}. The analogs of $\tau$ and $\Omega$ are, up to scaling factors, the difference of positions and the sum of momenta of the two particles. It is well known that in general the quantum theory allows one to define these two observables arbitrarily precisely, which is best illustrated by the celebrated Einstein-Podolsky-Rosen paradox \cite{EPR}. However, if the composite state is separable, i.e.\ it is a statistical mixture of well defined quantum states for individual subsystems, then the product of these two uncertainties has a lower bound derived by Tan \cite{TanPRA99}. Expressed in terms of time and frequency variables, this bound takes the form
\begin{equation}
\label{Eq:Separability}
\langle (\Delta \tau)^2 \rangle \langle (\Delta \Omega)^2 \rangle \ge {\textstyle 1}.
\end{equation}
Inserting this result into Eq.~(\ref{Eq:Deltatausym}) yields:
\begin{equation}
\label{Eq:NewInequality}
\langle (\Delta \tau' )^2 \rangle_\text{sym} \ge \langle (\Delta \tau)^2 \rangle +
\frac{(2 \beta L)^2}{\langle (\Delta \tau)^2 \rangle}.
\end{equation}
The above inequality is the central result of this paper. It defines the minimum broadening of temporal correlations between two light beams during propagation through dispersive media with opposite dispersion signs if no entanglement is present. A violation of this inequality is an unambiguous signature that the two beams have been initially prepared in an entangled state and it cannot be obtained within classical theory of optical radiation.

\section{Gaussian states}
\label{Sec:Gaussian}

The effect of nonlocal dispersion cancellation has been originally discussed for photon pairs generated through a decay of pump photons in the process of spontaneous parametric down-conversion \cite{FranPRA92}. For a monochromatic pump, energy conservation means that the frequencies of the twin photons must sum up to the frequency of the parent pump photon, implying that indeed $\langle (\Delta \Omega)^2 \rangle =0$. Consequently, temporal correlations are preserved. Such photon pairs can be considered as a postselected ensemble of the complete twin-beam state produced in spontaneous parametric down conversion. In order to expose the limitations of the classical theory, it is insightful to discuss the Wigner picture of a class of time-stationary Gaussian states \cite{ShapPRA10,ShapSunJOSAB94} that includes the twin-beam state as a particular case. This class is characterized by two positive spectra of phase-insensitive autocorrelation functions
\begin{equation}
S_{j}(\omega) = \int d\tau \, e^{-i\omega\tau} \,
 \Tr[\hat{\varrho}
  \hat{E}^{(-)}_j (t+\tau)
  \hat{E}^{(+)}_j (t)
 ], \quad j=1,2
\end{equation}
and the complex spectrum of the phase-sensitive cross-correlation function
\begin{equation}
S_{12}^{(p)}(\omega) = \int d\tau \, e^{i\omega\tau}
 \Tr[\hat{\varrho}
  \hat{E}^{(+)}_1 (t+\tau)
  \hat{E}^{(+)}_2 (t)
 ]
\end{equation}
while all the field means and other second-order correlation functions vanish.
The functions $S_{j}(\omega)$ can be interpreted as spectral intensity distributions per
unit time for individual beams \cite{MandelWolf},
\begin{equation}
\Tr[\hat{\varrho} \hat{\cal E}_j^\dagger (\omega) \hat{\cal E}_j (\omega') ]
= 2\pi S_{j}(\omega)\delta(\omega-\omega').
\end{equation}
For low intensities, $S_{12}^{(p)}(\omega)$ yields the probability amplitude of generating a pair of photons with frequencies $\omega$ and $-\omega$ per unit time,
\begin{equation}
\Tr[\hat{\varrho} \hat{\cal E}_1 (\omega) \hat{\cal E}_2 (\omega') ]
= 2\pi S_{12}^{(p)}(\omega)\delta(\omega+\omega').
\end{equation}
The Wigner function for the Gaussian states under consideration can be easily found with the help of Wick's theorem \cite{Wick} to take the form:
\begin{widetext}
\begin{equation}
\label{Eq:WignerGaussian}
W(t_1, \omega_1; t_2, \omega_2) \propto S_{1}(\omega_1) S_{2}(\omega_2)
+  \delta(\omega_1 + \omega_2) \int d\nu \, e^{i\nu(t_1-t_2)}
\bigl[S_{12}^{(p)}\bigl( {\textstyle\frac{1}{2}}(\omega_1 - \omega_2 + \nu)\bigr)\bigr]^\ast
S_{12}^{(p)}\bigl( {\textstyle\frac{1}{2}}(\omega_1 - \omega_2 - \nu)\bigr).
\end{equation}
\end{widetext}
The first term is simply the product of individual spectra, while the second one, confined to the phase space region where $\omega_1+\omega_2=0$, resembles a Wigner function for the two-photon probability amplitude $S_{12}^{(p)}(\omega)$ parameterized with $t_1-t_2$ and $(\omega_1-\omega_2)/2$.

The difference between predictions of the classical and the quantum theories lies in the relative magnitude of the two terms that form the Wigner function. A simple and general way to find relevant relations is to define an operator
\begin{equation}
\hat{\cal F}(\lambda)\! =\! \int_{-\epsilon_1}^{\epsilon_2} d\omega [\hat{\cal E}_1(\omega) + \lambda \hat{\cal E}_2^\dagger(-\omega)]
\end{equation} 
where $\epsilon_1,\epsilon_2 >0$  and $\lambda$ is a complex number, and to consider the necessarily nonnegative expectation value
$\Tr [\hat{\varrho}\hat{\cal F}(\lambda)\hat{\cal F}^\dagger(\lambda)]\ge 0$. Making use of the commutation relations given in Eq.~(\ref{Eq:CommSpectr}) and performing frequency integrals yields
\begin{equation}
1 + S_{1}(\omega) + 2\text{Re} [\lambda^\ast S_{12}^{(p)}(\omega)] +
|\lambda|^2 S_{2}(- \omega) \ge 0
\end{equation}
which holds for an arbitrary complex $\lambda$. This implies that \cite{ShapSunJOSAB94}:
\begin{equation}
\label{Eq:S12ineq}
|S_{12}^{(p)}(\omega)|^2 \le [1 + S_1(\omega)] S_2(-\omega).
\end{equation}

The inequality derived in Eq.~(\ref{Eq:S12ineq}) shows that in the limit of low intensities the marginal spectra $S_1(\omega_1)$ and $S_2(\omega_2)$ are allowed to scale as the square of the two-photon probability amplitude. In this case the product $S_1(\omega_1)S_2(\omega_2)$ appearing in the Wigner function calculated in Eq.~(\ref{Eq:WignerGaussian}) is fourth order in $S_{12}^{(p)}(\omega)$. It can therefore be neglected compared to the second term, which exhibits strong time-frequency correlations that are behind the effect of nonlocal dispersion cancellation and the violation of the inequality (\ref{Eq:NewInequality}).

For classical fields, one needs to consider statistical averages involving stochastic fields ${\cal E}_1$ and ${\cal E}_2$ and repeat steps leading to Eq.~(\ref{Eq:S12ineq}). However, the commutativity of the classical fields implies that $|S_{12}^{(p)}(\omega)|^2 \le S_1(\omega) S_2(-\omega)$. This means that the product of the marginal spectra $S_1(\omega_1)S_2(\omega_2)$ scales at least quadratically with $S_{12}^{(p)}(\omega)$ and contributes to the Wigner function with at best the same magnitude as the second term. As a result, the temporal distribution of detection events has a constant background shown schematically in Fig.~\ref{Fig:whatistau}(b), which severely affects the time difference variance. For time stationary fields it makes $\langle (\Delta \tau )^2 \rangle$ diverge to infinity. If the source emission time is restricted, for example with shutters opened for a finite period, $\langle (\Delta \tau )^2 \rangle$ would be entirely dominated by the constant background. In either case,
the inequality (\ref{Eq:NewInequality}) would be satisfied as expected for classical fields. Thus the use of time difference variance as a quantitative measure of temporal correlations allows one to draw a clear distinction between quantum and classical models of nonlocal dispersion cancellation \cite{ShapPRA10,FranPRA10}.

\section{Experimental considerations}
\label{Sec:Experimental}

The results of the experiment conducted by Baek et al. \cite{BaekChoOPEX09} on nonlocal dispersion cancelation are inconclusive with respect to our criterion, as the reduction in the time difference variance has been demonstrated only against the case with the negative-dispersion medium absent in one arm of the setup. To test the inequality (\ref{Eq:NewInequality}), one needs to measure time difference variance before and after dispersive propagation, the latter in a symmetrized scenario, and to determine independently effective dispersion. A practical issue in an experimental test may be the jitter of single photon detectors that limits the timing resolution. In order to assess its impact, let us assume that the intrinsic source variance $\langle(\Delta \tau)^2\rangle_{\text{source}}$ is enhanced by an additive term $\JitterVariance$ describing the temporal uncertainty of the detector response, yielding the actually observed time difference variance
$\langle(\Delta \tau)^2\rangle_{\text{obs}} = \langle(\Delta \tau)^2\rangle_{\text{source}} + \JitterVariance$. For down-conversion sources the intrinsic variance $\langle(\Delta \tau)^2\rangle_{\text{source}}$ depends primarily on the bandwidths of the spectra of individual beams, and for broadband down-conversion it can be made well below $(100~\text{fs})^2$ \cite{Broadband}. In this regime, $\JitterVariance$ becomes the dominant contribution to the observed time difference variance, as the timing resolution of currently available single-photon detectors is of the order of $50~\text{ps}$ \cite{TimingResolution}.

A prerequisite to observe a signature of entanglement is to maintain the violation of the separability criterion given in Eq.~(\ref{Eq:Separability}) even when the temporal correlations are affected by the jitter. This gives a condition for the sum $\Omega$ of the frequencies of the down-converted photons in the form $\langle (\Delta\Omega)^2 \rangle \ll [\JitterVariance]^{-1}$, which is effectively a constraint on the spectral linewidth of the cw laser pumping the nonlinear medium. This regime can be reached using a narrow linewidth pump laser. Further, in the inequality (\ref{Eq:NewInequality}) both
$\langle (\Delta \tau' )^2 \rangle_\text{sym}$ on the left-hand side and $\langle (\Delta \tau)^2 \rangle$ are equally affected by the jitter variance, while the dispersion-induced term ${(2 \beta L)^2}/{\langle (\Delta \tau)^2 \rangle}$ decreases when the denominator is enhanced by $\JitterVariance$. This makes it harder to violate an inequality calculated for actually observed $\langle (\Delta \tau' )^2 \rangle_\text{sym obs}$ and $\langle (\Delta \tau)^2 \rangle_\text{obs}$ compared to that with variances free from jitter effects. The remaining issue is the magnitude of the dispersion-induced term compared to the time difference variance. To make a statistically significant observation, we would like the dispersion term to be non-negligible to the initial time difference, i.e.\ ${(2 \beta L)^2}/{\langle (\Delta \tau)^2 \rangle_{\text{obs}}} \approx \langle (\Delta \tau)^2 \rangle_{\text{obs}}$. If the observed time difference variance is dominated by the detector jitter, this leads to a condition $2\beta L \approx \JitterVariance$. This condition is more challenging to fulfill, as in the exemplary experiment of Baek {\em et al.} \cite{BaekChoOPEX09} we had $ 2\beta L \approx (8~\text{ps})^2$, which is substantially smaller than typical $\JitterVariance$. This relation could be improved by transmitting fields through engineered highly dispersive optical fibers \cite{DispersiveFibers} and by using detectors based on parametric upconversion \cite{UpconvertingDetectors} that enable precise temporal gating by combining the signal with a short auxiliary pulse. Overall, a careful choice of the experimental regime should enable a violation of the inequality (\ref{Eq:NewInequality}).

\section{Conclusions}
\label{Sec:Conclusions}

In conclusion, we presented an inequality that reveals the role of entanglement in nonlocal dispersion cancellation. It is worth noting that an analogous discussion can be carried out for the spatial degree of freedom when two light beams are subjected to suitably arranged diffractive propagation \cite{KoelIEEE94}. Another interesting question would be to analyze the time interval statistics \cite{BaraBlakPR80}, whose variance could remain finite even in the presence of uniform background of detection events that occurs in the classical case analyzed above.

\section*{ACKNOWLEDGMENTS}

We acknowledge insightful discussions with R. Demkowicz-Dobrza\'{n}ski, I. A. Walmsley, and P. Wasylczyk. This work was supported by the Foundation for Polish Science TEAM project co-financed by the EU European Regional Development Fund.

\pagebreak

\end{document}